\documentstyle{article}
\begin{document}
\title{van Vleck determinants:\\
       geodesic focussing and defocussing\\
       in Lorentzian spacetimes.}
\author{Matt Visser\cite{e-mail}\\
        Physics Department\\
        Washington University\\
        St. Louis\\
        Missouri 63130-4899}

\date{November 1992}
\maketitle

\begin{abstract}
The van Vleck determinant is an ubiquitous object, arising in many
physically interesting situations such as: (1) WKB approximations
to quantum time evolution operators and Green functions. (2)
Adiabatic approximations to heat kernels. (3) One loop approximations
to functional integrals. (4) The theory of caustics in geometrical
optics and ultrasonics. (5) The focussing and defocussing of geodesic
flows in Riemannian manifolds.

While all of these topics are interrelated, the present paper is
particularly concerned with the last case and presents extensive
theoretical developments that aid in the computation of the van
Vleck determinant associated with geodesic flows in Lorentzian
spacetimes.  {\sl A fortiori} these developments have important
implications for the entire array of topics indicated.


To appear in Physical Review D47 (1993) 15 March.
\end{abstract}
\newpage
\section{INTRODUCTION}

The van Vleck determinant is a truly ubiquitous object in mathematical
physics. Introduced by van Vleck in 1928 \cite{vanVleck}, it was
first utilized  in elucidating the nature of the classical limit
of quantum mechanics via WKB techniques. Further developments were
due to Morette \cite{Morette}. A nice discussion of this original
line of development can be found in Pauli's lecture notes \cite{Pauli}.

In a different vein, de Witt \cite{deWitt-I,deWitt-II,deWitt-III,deWitt-IV}
and others have developed extensive and powerful adiabatic expansion
methods for approximating quantum Green functions in Lorentzian
and Riemannian manifolds.  Textbook discussions may be found in
Birrell and Davies \cite{Birrell-Davies} and Fulling \cite{Fulling}.

By extension, these adiabatic techniques lead to powerful
point--splitting techniques. These techniques are useful, for
example, for estimating the vacuum expectation value of the
renormalized stress--energy tensor of a quantum field propagating
on a Lorentzian spacetime. As will soon be apparent, in this
particular context the van Vleck determinant is essentially a
measure of the tidal focussing and/or defocussing of geodesic flows
in spacetime.

This paper will develop several formal techniques for getting a
handle on the van Vleck determinant. The evolution of the van Vleck
determinant as one moves along a geodesic will be explicitly
evaluated in terms of the expansion of the geodesic flow. The
relationship with the (spacelike/lightlike/timelike) versions of
the Raychaudhuri equation is thus made manifest. Constraints, such
as the Weak Energy Condition (WEC), are then used to develop powerful
inequalities limiting the behaviour of  the van Vleck determinant.
The evolution of the van Vleck determinant is then reformulated in
terms of tidal effects --- the evolution of a set of Jacobi fields
under the influence of the full Riemann tensor.  A formal solution
to the resulting integral equation is presented.  Using these formal
techniques, a weak field approximation is developed. This weak
field approximation will be seen to depend only on the Ricci tensor
of spacetime.  In a similar vein, a short distance approximation
is developed. Finally, an asymptotic estimate is provided of the
behaviour of the determinant as one  moves out to infinity.

The physical problem that originally stimulated my interest in
these  investigations was Hawking's Chronology Protection Conjecture
\cite{Hawking-I,Hawking-II}.  That conjecture led to several
different computations of the vacuum expectation value of the
renormalized stress--energy tensor for quantum fields propagating
in spacetimes on the verge of violating chronology protection
\cite{Kim-Thorne,Klinkhammer,CPC,Structure}. The van Vleck determinant
is an overall prefactor occurring in all those computations.

It is nevertheless true that much of the mathematical machinery
developed in this paper has a considerably wider arena of applicability.
Accordingly, some effort will be made to keep  the discussion as
general as reasonably possible.

\underbar{Notation:} Adopt units where $c\equiv 1$, but all other
quantities retain their usual dimensionalities. In particular
$G=\hbar/m_P^2 = \ell_P^2/\hbar$. The metric signature is taken to
be $(-,+,\cdots,+)$. General conventions follow Misner, Thorne,
and Wheeler \cite{MTW}.

\section{THE VAN VLECK DETERMINANT}

\subsection{General definition}

Consider an arbitrary mechanical system, of $n$ degrees of freedom,
governed by a Lagrangian ${\cal L}(\dot q, q)$. Solve the equations
of motion to find the (possibly not unique) path $\gamma$ passing
through the points $(q_i,t_i)$ and $(q_f,t_f)$. Then calculate the
action of that path
\begin{equation}
S_\gamma(q_i,t_i;q_f,t_f) \equiv \int_\gamma {\cal L}(\dot q, q) dt.
\end{equation}
The van Vleck determinant is then defined as
\begin{equation}
\Delta_\gamma(q_i,t_i;q_f,t_f) \equiv
(-1)^n \det\left\{
{\partial^2 S_\gamma(q_i,t_i;q_f,t_f)
\over\partial q_i \, \partial q_f}
\right\}_.
\end{equation}
The van Vleck determinant, in this original incarnation, occurs as
a prefactor in the WKB approximation to the quantum time evolution
operator \cite{vanVleck,Morette,Pauli}. In the Schrodinger picture
\begin{eqnarray}
&&<q_f|\exp\{-iH(t_f-t_i)/\hbar\}|q_i> \approx  \nonumber\\
&&\qquad (2\pi i\hbar)^{-n/2} \; \sum_\gamma
\sqrt{\Delta_\gamma(q_i,t_i;q_f,t_f)}
\exp\{+iS_\gamma(q_i,t_i;q_f,t_f)/\hbar\}.
\end{eqnarray}
In the particular case of a geodesic flow on a Lorentzian or
Riemannian manifold the appropriate action to be inserted in this
definition is the geodetic interval between a pair of points.

\subsection{Geodetic interval}

Consider a Lorentzian spacetime of total dimensionality $(d+1)$.
That is: $d$ space dimensions, $1$ time dimension. With appropriate
changes, these results apply also to a Riemannian manifold of total
dimensionality $(d+1)$. To go from the Lorentzian to the Riemannian
case the only significant change is that one should forget all the
subtleties associated with lightlike (null) geodesics.

The geodetic interval may be defined by:
\begin{equation}
\sigma_\gamma(x,y) \equiv \pm{1\over2} [s_\gamma(x,y)]^2_.
\end{equation}
Here we take the upper ($+$) sign if the geodesic $\gamma$ from
the point $x$ to the point $y$ is spacelike.  We take the lower
($-$) sign if this geodesic $\gamma$ is timelike. In either case
we define the geodesic distance $s_\gamma(x,y)$ to be positive
semi-definite.

Note that, provided the geodesic from $x$ to $y$ is not lightlike,
\begin{eqnarray}
\nabla^x_\mu \sigma_\gamma(x,y)
&=& \pm s_\gamma(x,y) \; \nabla^x_\mu s_\gamma(x,y), \\
&=&  + s_\gamma(x,y) \; t_\mu(x;\gamma; x \leftarrow y),\\
&=&  + \sqrt{2 |\sigma_\gamma(x,y)|}\;\; t_\mu(x \leftarrow y).
\end{eqnarray}
Here $t^\mu(x; \gamma; x \leftarrow y) \equiv \pm g^{\mu\nu}
\nabla^x_\nu s_\gamma(x,y)$ denotes the unit tangent vector at the
point $x$ pointing along the geodesic $\gamma$ away from the point
$y$. When no confusion results we may abbreviate this by $t^\mu(x
\leftarrow y)$ or even $t^\mu(x)$.

If the geodesic from $x$ to $y$ is lightlike things are somewhat
messier. One easily sees that for lightlike geodesics $\nabla^x_\mu
\sigma_\gamma(x,y)$ is a null vector. To proceed further one must
introduce a canonical observer, described by a unit timelike vector
$V^\mu$ at the point $x$.  By parallel transporting this canonical
observer along the geodesic one can set up a canonical frame that
picks out a particular canonical affine parameter:
\begin{eqnarray}
&&\nabla^x_\mu \sigma_\gamma(x,y) =
+\zeta_\gamma(x,y) \; l_\mu(x; \gamma; x \leftarrow y), \\
&&l_\mu(x; \gamma; x \leftarrow y)\; V^\mu = -1, \\
&&\zeta_\gamma(x,y) = - V^\mu \nabla^x_\mu \; \sigma_\gamma(x,y).
\end{eqnarray}
Note that this affine parameter $\zeta$ can, crudely, be thought
of as a distance along the null geodesic as measured by an observer
with four--velocity $V^\mu$. By combining the null vector $l^\mu$
with the timelike vector $V^\mu$, one can construct a second
canonical null vector: $m^\mu \equiv 2V^\mu - l^\mu$. A quick
calculation shows
\begin{eqnarray}
&&\nabla^x_\mu \zeta_\gamma(x,y) =
- {1\over2} \; m_\mu(x; \gamma; x \leftarrow y), \\
&&m_\mu(x) \; V^\mu = -1; \\
&&l_\mu(x) \; m_\mu(x) = -2.
\end{eqnarray}
Unfortunately, while spacelike and timelike geodesics can be treated
in a unified formalism as alluded to above, the subtleties involved
with lightlike geodesics will require the presentation of several
tedious variations on the general analysis.

\subsection{Specific definition: geodesic flow}

Consider geodesic flow in a Lorentzian spacetime.  In the present
context the (scalarized) van Vleck determinant is defined by
\begin{equation}
\Delta_\gamma(x,y) \equiv (-1)^d \;
   {\det\{\nabla^x_\mu \nabla^y_\nu \sigma_\gamma(x,y)\}
    \over \sqrt{g(x) g(y)} }.
\end{equation}
This definition has a nice interpretation in terms of the Jacobian
associated with the change of variables from $(x,t)$ to $(x,y)$.
One may specify a geodesic either by (a) specifying a single point
on the geodesic, $x$, and the tangent vector, $t$, at that point;
or by (b) specifying two separate points, $(x,y)$, on the geodesic.
The (scalarized) Jacobian associated with this change of variables
is
\begin{equation}
J_\gamma(x,y)
\equiv
{ \det\left\{ \partial(x,t)/\partial(x,y) \right\}
  \over \sqrt{g(x) g(y)} }
=
{  {\det}'\left\{{\partial t/\partial y}\right\}
   \over \sqrt{g(x) g(y)} }.
\end{equation}
Here the $\det'$ indicates the fact that we should ignore the known
trivial zero eigenvalue(s) in determining this Jacobian.

If the points $x$ and $y$ are spacelike or timelike separated then
the one trivial zero arises from the fact that the tangent vector
is normalized, $t^\mu t_\mu = \pm1$, so that $t^\mu(x) \nabla^y_\nu
t_\mu(x) =0$.  To see the connection with the van Vleck determinant,
observe
\begin{eqnarray}
\nabla^x_\mu \nabla^y_\nu \sigma_\gamma(x,y)
&=&
\nabla^y_\nu \left[ s_\gamma(x,y) \; t_\mu(x;\gamma; x \leftarrow y) \right],
\nonumber \\
&=&
s_\gamma(x,y) \;\nabla^y_\nu t_\mu(x;\gamma; x \leftarrow y)
\pm t_\nu(y) \; t_\mu(x).
\end{eqnarray}
By adopting suitable coordinates at $x$, and independently at $y$,
one can make the tangent vectors $t_\mu(x)$ and $t_\mu(y)$ both
lie in the ``$t$'' direction (if they are timelike) or  both lie
in the ``$z$'' direction (if they are spacelike). Then by inspection
\begin{eqnarray}
\Delta_\gamma(x,y) &=&  \pm[-s_\gamma(x,y)]^d \;
   {\det'\{\nabla^y_\nu t_\mu(x;\gamma; x \leftarrow y) \}
    \over \sqrt{g(x) g(y)} }  \nonumber\\
    &=& \pm[-s_\gamma(x,y)]^d  \; J_\gamma(x,y).
\end{eqnarray}

If the points $x$ and $y$ are lightlike separated then the two
trivial zeros arise from the fact that the tangent vector $l^\mu$
satisfies two constraints: $l^\mu l_\mu = 0$, and $l^\mu V_\mu =
-1$. Thus both $l^\mu(x) \nabla^y_\nu t_\mu(x) =0$, and $V^\mu(x)
\nabla^y_\nu t_\mu(x) =0$.  To precisely determine the connection
with the van Vleck determinant requires some subtlety. For a point
$x$ close to, but not quite on, a null geodesic emanating from $y$
one may usefully decompose the gradient of the geodetic interval
as
\begin{equation}
\nabla^x_\mu \sigma_\gamma(x,y) =
+\zeta_\gamma(x,y) \; l_\mu(x; \gamma; x \leftarrow y)
+\xi_\gamma(x,y) \; m_\mu(x; \gamma; x \leftarrow y).
\label{affine}
\end{equation}
Here $\zeta$ and $\xi$ are to be thought of as curvilinear null
coordinates. They are the curved space generalizations of $(t\pm
x)/2$. One is now in a position to calculate
\begin{eqnarray}
\nabla^x_\mu \nabla^y_\nu \sigma_\gamma(x,y)
&=&
\nabla^y_\nu \left[
\zeta(x,y) \; l_\mu(x \leftarrow y)
+\xi(x,y) \; m_\mu(x \leftarrow y)
 \right],
\nonumber \\
&=&
\zeta(x,y) \;\nabla^y_\nu l_\mu(x;\gamma; x \leftarrow y)
+\xi(x,y) \;\nabla^y_\nu m_\mu(x;\gamma; x \leftarrow y)
\nonumber \\
&&\qquad - {1\over2} m_\nu(y) \; l_\mu(x)
         - {1\over2} l_\nu(y) \; m_\mu(x).
\end{eqnarray}
Now go to the light cone, by setting $\xi=0$.  By adopting suitable
coordinates at $x$ one can arrange: $l_\mu(x) = (1,1,0,0)$ and
$m_\mu(x) = (1,-1,0,0)$. Independently, one can arrange the same
to be true at $y$. Finally, careful  inspection of the above reveals
\begin{eqnarray}
\Delta_\gamma(x,y)
&=&  (-1)^d [\zeta_\gamma(x,y)]^{(d-1)} \;
   {\det'\{\nabla^y_\nu l_\mu(x;\gamma; x \leftarrow y) \}
    \over \sqrt{g(x) g(y)} }  \nonumber\\
&=& (-1)^d [\zeta_\gamma(x,y)]^{(d-1)}  \; J_\gamma(x,y).
\end{eqnarray}

\subsection{Elementary results}

In view of the equation
\begin{equation}
\nabla^x_\mu \sigma_\gamma(x,y)
= s_\gamma(x,y) \; t_\mu(x;\gamma; x \leftarrow y),
\end{equation}
one has as an exact result (valid also on the light cone)
\begin{equation}
\sigma_\gamma(x,y) =
 {1\over2} g^{\mu\nu} \; \nabla^x_\mu \sigma_\gamma(x,y)
                      \; \nabla^x_\nu \sigma_\gamma(x,y).
\end{equation}
Repeated differentiations and contractions result in
\cite{deWitt-I,deWitt-II,deWitt-III,deWitt-IV}
\begin{equation}
\nabla_x^\mu \left[ \Delta_\gamma(x,y) \;
\nabla^x_\mu\sigma_\gamma(x,y) \right]
= (d+1) \Delta_\gamma(x,y).
\label{master}
\end{equation}
Assuming that $x$ and $y$ are either timelike separated or spacelike
separated, this may be rewritten as follows
\begin{equation}
\nabla_x^\mu \left[ \Delta_\gamma(x,y) \; s_\gamma(x,y) \; t_\mu(x) \right]
= (d+1) \Delta_\gamma(x,y).
\end{equation}
Use of the Leibnitz rule leads to
\begin{eqnarray}
&& s_\gamma(x,y) \; t_\mu(x) \; \nabla_x^\mu \Delta_\gamma(x,y)
\nonumber   \\
&& \pm  t_\mu(x) \; t^\mu(x) \; \Delta_\gamma(x,y)
\nonumber   \\
&&+ s_\gamma(x,y)\; \Delta_\gamma(x,y) \; \nabla_x{}^\mu t_\mu(x)
\nonumber   \\
&&\qquad =  (d+1) \Delta_\gamma(x,y).
\end{eqnarray}
One notes that $t^\mu(x)$ defines a normalised spacelike or timelike
vector field. Thus
\begin{equation}
t_\mu(x) \; \nabla_x^\mu \Delta_\gamma(x,y)
= \left( {d\over s_\gamma(x,y)} - (\nabla_x{}^\mu t_\mu(x)) \right)
\Delta_\gamma(x,y).
\end{equation}
Now define $\Delta_\gamma(s)$ to be the van Vleck determinant
calculated at a proper distance $s$ along the geodesic $\gamma$ in
the direction away from $y$ and towards $x$. Note the boundary
condition that $\Delta_\gamma(s=0) \equiv 1$; and define
\begin{equation}
t^\mu(x) \nabla^x_\mu  f \equiv {df\over ds}.
\end{equation}
One finally obtains a first order differential equation governing
the evolution of the van Vleck determinant
\begin{equation}
{d\Delta_\gamma(s)\over ds}
= \left( {d\over s} -   \theta \right) \Delta_\gamma(s).
\end{equation}
Here $\theta$ is the expansion of the geodesic spray defined by
the integral curves of $t^\mu(x)$, that is, $\theta \equiv \nabla_\mu
t^\mu$.  Direct integration yields
\begin{equation}
\Delta_\gamma(x,y) = s_\gamma(x,y)^d
\exp\left( - \int_\gamma \theta ds \right)_.
\end{equation}
Here the integration is to be taken along the geodesic $\gamma$
from the point $y$ to the point $x$. The interpretation of this
result is straightforward: take a geodesic spray of trajectories
emanating from the point $y$. If they were propagating in flat
space, then after a proper distance $s$ the (relative) transverse
density of trajectories would have fallen to $s^{-d}$. Since they
are not propagating in flat spacetime the actual (relative) transverse
density of trajectories is given by the exponential of minus the
integrated expansion. The van Vleck determinant is then the ratio
between the actual density of trajectories and the anticipated flat
space result. Note that by explicit calculation, the van Vleck
determinant is completely symmetric in $x$ and $y$.

An analogous development holds if the points $x$ and $y$
are connected by a null geodesic. The details are, as one has by
now grown to expect, somewhat tedious.  One returns to equation
(\ref{master}) and uses equation (\ref{affine}) to derive
\begin{equation}
\nabla_x^\mu \left[ \Delta_\gamma(x,y)
\{ \zeta \; l_\mu(x) + \xi \; m_\mu(x) \}\right]
= (d+1) \Delta_\gamma(x,y).
\end{equation}
Use of the Leibnitz rule, followed by the limit $\xi\to 0$,  now
leads to
\begin{eqnarray}
&& \zeta \; l_\mu(x) \; \nabla_x^\mu \Delta_\gamma(x,y)
\nonumber   \\
&&- l_\mu(x) \; m^\mu(x) \; \Delta_\gamma(x,y)
\nonumber   \\
&&+ \zeta \; \Delta_\gamma(x,y) \; \nabla_x{}^\mu l_\mu(x)
\nonumber   \\
&&\qquad =  (d+1) \Delta_\gamma(x,y).
\end{eqnarray}
One recalls that, by definition, $l^\mu(x)  m_\mu(x)= -2$.  Adopting
suitable definitions, analogous to the non--null case, one derives
\begin{equation}
{d\Delta_\gamma(\zeta)\over d\zeta}
= \left( {[d-1]\over \zeta} -  \hat\theta \right) \Delta_\gamma(\zeta).
\label{null-evolution}
\end{equation}
Direct integration yields
\begin{equation}
\Delta_\gamma(x,y) = \zeta_\gamma(x,y)^{[d-1]}
\exp\left( - \int_\gamma \hat\theta d\zeta \right)_.
\end{equation}
In particular, note that while the affine parameter $\zeta$, and
the expansion $\hat\theta \equiv \nabla_\mu l^\mu$ both depend on
the canonical observer $V^\mu$, the overall combination is independent
of this choice. The appearance of the exponent $d-1$ for null
geodesics, as contrasted to the exponent $d$ for spacelike or
timelike geodesics might at first be somewhat surprising. The
ultimate reason for this is that in $(d+1)$ dimensional spacetime
the set of null geodesics emanating from a point $x$ sweeps out a
$d$ dimensional submanifold. As one moves away from $x$ an affine
distance $\zeta$ into this submanifold the relative density of null
geodesics falls as $\zeta^{1-d}$. In $(3+1)$ dimensional spacetime,
this is just the inverse square law for luminosity.

Equation (\ref{null-evolution}) above should be compared with
equation (22) of Kim and Thorne \cite{Kim-Thorne}. The typographical
error in that equation (the $3$ should be a $2$) fortunately does
not propagate into the rest of their paper. The virtue of this
tedious but elementary analysis is that one now has a unified
framework applicable in all generality  to cases of timelike,
lightlike, or spacelike separated points $x$ and $y$.

\subsection{Inequalities}

Now that one has these general formulae for the van Vleck determinant,
powerful inequalities can be derived by applying the Raychaudhuri
equation and imposing suitable convergence conditions. Restricting
attention to vorticity free (spacelike or timelike) geodesic flows
the Raychaudhuri equation particularizes to
\begin{equation}
{d\theta\over ds} =
- \left( R_{\mu\nu} t^\mu t^\nu \right)
-2\sigma^2 -{\theta^2\over d}.
\end{equation}
For a concrete textbook reference, this is a special case of equation
(4.26) of Hawking and Ellis \cite{Hawking-Ellis}, generalized to
arbitrary dimensionality.  While Hawking and Ellis are discussing
timelike geodesics, the choice of notation in this paper guarantees
that the discussion can be carried over to spacelike geodesics
without alteration.  The quantity $\sigma$ denotes the shear of
the geodesic congruence, and $\sigma^2$ is guaranteed to be positive
semi--definite.

\noindent\underbar{Definition:} A Lorentzian spacetime is said to
satisfy the (timelike/null/spacelike) convergence condition if for
all (timelike/null/spacelike) vectors $t^\mu$: $\left( R_{\mu\nu}
t^\mu t^\nu \right) \geq 0$.
\vrule height 0.9 ex width 0.8 ex depth -0.1 ex

\noindent\underbar{Definition:} A Riemannian manifold is said to
possess semi--positive Ricci curvature if for all vectors
$t^\mu$:  $\left( R_{\mu\nu} t^\mu t^\nu \right) \geq 0$.
\vrule height 0.9 ex width 0.8 ex depth -0.1 ex

If the points $x$ and $y$ are (timelike/spacelike) separated, and
the spacetime satisfies the (timelike/spacelike) convergence
condition then one has the inequality
\begin{equation}
{d\theta\over ds} \geq -{\theta^2\over d}.
\end{equation}
This inequality is immediately integrable
\begin{equation}
{1\over\theta(s)} \geq {1\over\theta(0)} + {s\over d}.
\end{equation}
For the geodesic spray under consideration, one has $\theta(s) \to
d/s$ as $s \to 0$. Therefore $[1/\theta(0)] = 0$. One has thus
derived an inequality, valid for geodesic sprays, under the assumed
convergence conditions
\begin{equation}
{\theta(s)}  \leq + {d\over s}.
\end{equation}
This implies that the van Vleck determinant is a monotone function
of arc length, $d\Delta/ds \geq 0$, and immediately integrates to
an inequality on the determinant itself $ \Delta_\gamma(x,y) \geq
1$.

For null geodesic flows the Raychaudhuri equation becomes
\begin{equation}
{d\hat\theta\over d\zeta} =
- \left( R_{\mu\nu} l^\mu l^\nu \right)
-2\sigma^2 -{{\hat\theta}^2\over(d-1)}.
\end{equation}
See equation (4.35) of Hawking and Ellis \cite{Hawking-Ellis}.  If
the spacetime now satisfies the null convergence condition one
can deduce the inequality
\begin{equation}
{d\hat\theta\over d\zeta} \geq -{\hat\theta^2\over d-1}.
\end{equation}
Upon integration, and use of the relevant boundary condition, one
derives in a straightforward manner the inequality
\begin{equation}
{\hat\theta(\zeta)}  \leq + {(d-1)\over \zeta}.
\end{equation}
Which implies that the van Vleck determinant is a monotone function
of the null affine parameter, $d\Delta/d\zeta \geq 0$, and immediately
integrates to an inequality on the determinant itself $ \Delta_\gamma(x,y)
\geq 1$.  One is now in a position to enunciate the following
results.

\noindent\underbar{Theorem:} In any Lorentzian spacetime, if the
points $x$ and $y$ are (timelike/null/ spacelike) separated, and
the spacetime satisfies the (timelike/null/spacelike) convergence
condition, then the van Vleck determinant is a monotone function
of affine parameter and is bounded from below: $\Delta_\gamma(x,y)
\geq 1$. \vrule height 0.9 ex width 0.8 ex depth -0.1 ex

\noindent\underbar{Theorem:} In any Riemannian manifold of
semi--positive Ricci curvature, the van Vleck determinant is a
monotone function of affine parameter and is bounded from below:
$\Delta_\gamma(x,y) \geq 1$.
\vrule height 0.9 ex width 0.8 ex depth -0.1 ex

To see the physical import of the (timelike/null/spacelike)
convergence conditions, consider a type I stress--energy tensor.
The cosmological constant, if present, is taken to be subsumed into
the definition of the stress--energy. Then $T_{\mu\nu} \sim {\rm
diag} [\rho; p_i]_{\mu\nu}$, where the $p_i$ are the $d$ principal
pressures. From the Einstein equations, $R_{\mu\nu} - {1\over2}R
g_{\mu\nu} = 8\pi G \; T_{\mu\nu}$, one deduces $R_{\mu\nu} \sim
4\pi G \; {\rm diag} [ \rho+\sum_j p_j; \rho+2p_i-\sum_j p_j]_{\mu\nu}$.

The timelike convergence condition implies: (1) $\rho+\sum_j
p_j \geq 0$; (2) $\forall i, \rho+ p_i \geq 0$. This is equivalent
to the Strong Energy Condition (SEC).

The null convergence condition implies:  $\forall i, \rho+
p_i \geq 0$. This is a somewhat weaker restriction than the Weak
Energy Condition (WEC). (The WEC would require the additional
constraint $\rho\geq 0$.)

The spacelike convergence condition implies: (1) $\forall i, \rho+
2 p_i - \sum_j p_j \geq 0$; (2) $ \forall i, \rho+ p_i \geq 0$.
This is a rather strong constraint that is not equivalent to any
of the standard energy conditions. Particularize to (3+1) dimensions,
then the constraints (1) can be made more explicit as:  $\rho +
p_1 - p_2 - p_3 \geq 0$; $\rho + p_2 - p_3 - p_1 \geq 0$; $\rho +
p_3 - p_1 - p_2 \geq 0$. By adding the first two of these constraints
one may deduce $\rho - p_3 \geq 0$. Similarly $\forall i, \rho-
p_i \geq 0$.  Thus the spacelike convergence condition implies the
Dominant Energy Condition (DEC), which in turn implies the Weak
Energy Condition (WEC). The implication does not hold in the reverse
direction. In particular, for an electromagnetic field of zero
Poynting flux $T_{\mu\nu} \sim \rho\cdot {\rm diag}
[+1;-1;+1;+1]_{\mu\nu}$. This stress--energy tensor satisfies the
Dominant Energy Condition but not the spacelike convergence condition.
A canonical example of a stress--energy tensor that \underbar{does}
satisfy the spacelike convergence condition is a perfect fluid that
satisfies the Dominant Energy Condition: $\rho + p \geq 0$; $\rho
- p \geq 0$.

Note that the timelike and spacelike convergence conditions separately
imply the null convergence condition.

The physical reason behind the inequality satisfied by the van
Vleck determinant is now manifest: ``Ordinary'' matter produces an
attractive gravitational field. An attractive gravitational field
focuses geodesics, so that they do not spread out as much as they
would in flat space. Then the van Vleck determinant is bounded from
below, and in fact continues to grow as one moves along any geodesic.

One should note however, that several interesting classes of
spacetimes violate one or more of these convergence conditions. This
makes the inequalities derived here somewhat less useful
than they might otherwise be, and suggests that it would be useful
to explore the possibility of deriving more general inequalities
based on weaker energy conditions.  For example, spacetimes containing
the interesting class of objects known as traversable wormholes
violate all three of these convergence conditions \cite{Morris-Thorne,MTY}.
Indeed in that particular class of spacetimes the required presence
of ``exotic'' matter leads to van Vleck determinants that are
arbitrarily close to zero \cite{Kim-Thorne}.

\subsection{Reformulation: Tidal focussing}

Direct computations using the preceding results are unfortunately
rather difficult. To get a better handle on the van Vleck determinant
we shall first apparently make the problem more complicated by
deriving a second order differential equation for the van Vleck
determinant.  By going to the second order formalism it is possible
to relate the van Vleck determinant directly to the tidal forces
induced by the full Riemann tensor --- more specifically to the focussing
and defocussing effects induced by tidal forces.

The most direct route to this end is to pick up a standard reference
such as Hawking and Ellis \cite{Hawking-Ellis}. Equation (4.20) on page 83
implies that
\begin{equation}
\int\theta ds = \ln\det[A^\mu{}_\nu(s)].
\end{equation}
Here $A^\mu{}_\nu(s)$ is the $d\times d$ matrix describing the
evolution of the separation of infinitesimally nearby geodesics.
This formalism may also be reformulated in terms of the Jacobi
fields associated with the geodesic $\gamma$.  Note that the $d\times
d$ matrix  $A^\mu{}_\nu(s)$  may equivalently be thought of as a
$(d+1)\times(d+1)$ matrix that is trivial in the extra entries:
$A_{(d+1)} = A_d\oplus s$.  In terms of $A$
\begin{equation}
\Delta_\gamma(x,y) = s_\gamma(x,y)^d \; \det(A_d^{-1})
                   = \det( s_\gamma(x,y) A^{-1} ).
\end{equation}
Where the last determinant can be taken in the sense of either a
$d\times d$ or a $(d+1)\times(d+1)$ matrix.

An alternative, more explicit, but also more tedious route to
establishing the preceding equation is to define the object
\begin{equation}
{\tilde A}(s) \equiv P \exp\{ \int_\gamma (\vec\nabla\otimes\vec t\,) ds \}.
\end{equation}
Here the symbol $P$ denotes the path ordering process. By definition
of path ordering
\begin{equation}
{d{\tilde A}\over ds} =  (\vec\nabla\otimes\vec t\,) \cdot
P \exp\{ -\int_\gamma (\vec\nabla\otimes\vec t\,) ds \}.
\end{equation}
On the other hand, ${\tilde A}(0)\equiv I$. Comparison with equation
(4.10) of Hawking and Ellis now shows that ${\tilde A}(s) \equiv
A(s)^\mu{}_\nu \; \partial_\mu\otimes dx^\nu$. Finally note that
\begin{equation}
\det{\tilde A} = \exp\{ \int_\gamma
                 \hbox{\rm tr}(\vec\nabla\otimes\vec t\,) ds \}
= \exp\{ \int_\gamma \theta ds \}.
\end{equation}
This reformulates the van Vleck determinant in terms of a path
ordered exponential
\begin{equation}
\Delta_\gamma(x,y) = s_\gamma(x,y)^d \;
\det\left[P \exp\{ -\int_\gamma (\vec\nabla\otimes\vec t\,) ds \}\right].
\end{equation}
This observation serves to illustrate formal similarities (and
differences)  between the van Vleck determinant and the Wilson loop
variables of gauge theories.

By taking a double derivative with respect to arc length the matrix
$A(s)$ may be shown to satisfy the second order differential equation
\begin{equation}
{d^2\over ds^2} A^\mu{}_\nu (s)
= - \left( R^\mu{}_{\alpha\sigma\beta} t^\alpha t^\beta \right)
A^\sigma{}_\nu.
\label{tidal}
\end{equation}
See Hawking and Ellis \cite{Hawking-Ellis}, equation (4.21) on page
83, particularized to a geodesic flow.
The boundary condition on $A(s)$ is that $A^\mu{}_\nu(s) \to s
\delta^\mu{}_\nu$ as $s\to0$, this being the flat space result for
a geodesic spray. This is the promised tidal formulation for the
van Vleck determinant.

A particularly pleasant feature of the tidal reformulation is that
the case of null geodesics can be handled without too much special
case fiddling. The analogue of the tidal equation is
\begin{equation}
{d^2\over d\zeta^2} {\hat A}^\mu{}_\nu (s)
= - \left( R^\mu{}_{\alpha\sigma\beta} \; l^\alpha l^\beta \right)
{\hat A}^\sigma{}_\nu.
\label{tidal-null}
\end{equation}
See Hawking and Ellis \cite{Hawking-Ellis}, equation (4.33) on page
88. Note that two of the eigenvectors of $\hat A$ suffer trivial
evolution. This is a consequence of the two normalization conditions
on the null tangent vector. Thus $\hat A$ can be thought of as
either a $(d+1) \times (d+1)$ matrix with two trivial entries, or
as a reduced $(d-1) \times (d-1)$  matrix. The formulations are
related by: ${\hat A}_{(d+1)} = {\hat A}_{(d-1)} \oplus \zeta I_2$.
Other results can be simply transcribed as needed.

\subsection{Reformulation: Formal solution}

To solve the above differential equation (in a formal sense),
introduce the one-dimensional retarded Green function
\begin{equation}
G_R(s_f,s_i) = \{s_f-s_i\} \Theta(s_f-s_i).
\end{equation}
Here $\Theta(s)$ denotes the Heavyside step function. For
notational convenience define
\begin{equation}
Q^\mu{}_\nu(s) = - \left( R^\mu{}_{\alpha\nu\beta} t^\alpha t^\beta \right).
\end{equation}

Integration of the second-order ``tidal'' equation for $\Delta_\gamma$,
with attention to the imposed boundary condition, leads to the
integral equation
\begin{equation}
A^\mu{}_\nu(s) =  s \delta^\mu{}_\nu
  + \int_0^s   G_R(s,s') Q^\mu{}_\sigma(s') A^\sigma{}_\nu(s') ds'.
\label{integral}
\end{equation}
More formally, one may suppress the explicit integration by regarding
$G_R(s,s')$, multiplication by $Q(s)$, and multiplication by $s$,
as functional operators. Then
\begin{equation}
(I - G_R Q) A  =  \{sI\}.
\end{equation}
This has the formal solution
\begin{eqnarray}
 A  &=&  (I - G_R Q)^{-1} \{sI\}      \nonumber\\
    &=&  \left(I + [G_R Q] + [G_R Q]^2 + \cdots \right) \{sI\}.
\label{formal}
\end{eqnarray}
This formal solution may also be derived directly from equation
(\ref{integral}) by continued iteration.  To understand what these
symbols mean, note that
\begin{equation}
[G_R Q]\{sI\} \equiv
\int_0^s G_R(s,s') [Q^\mu{}_\nu(s')] s' ds'
= \int_0^s (s-s')[ Q^\mu{}_\nu(s')] s' ds'.
\end{equation}
Similarly
\begin{equation}
[G_R Q]^2\{sI\} =
\int_0^s ds' \int_o^{s'}  ds''
(s-s') [Q^\mu{}_\alpha(s')] (s'-s'') [Q^\alpha{}_\nu(s'')] s''.
\end{equation}

The formal solution in terms of continued iteration is particularly
advantageous in the case where matter is concentrated on thin shells
\cite{Lanczos,Sen,Israel}. In that case the matrix $[Q^\mu{}_\nu(s)]$
is described by a series of delta functions. Due to the presence
of the Heavyside step function, the formal expansion terminates in
a finite number of steps
\begin{equation}
 A =  \left(I + [G_R Q] + [G_R Q]^2 + \cdots + [G_R Q]^N \right) \{sI\}.
\end{equation}
Here $N$ denotes the total number of shell--crossings. This type
of calculation will be explored in considerable detail in a subsequent
publication. For the time being, one
may observe from the above, that when the Riemann curvature is
concentrated on thin shells, the matrix $A$ is a piecewise linear
continuous matrix function of arc length. Consequently the reciprocal
of the van Vleck determinant $\Delta_\gamma(s)^{-1} = \det\{A/s\}$
is piecewise a Laurent polynomial in arc length.

\subsection{Weak field approximation}

These formal manipulations permit the derivation of a very nice
weak field approximation to the van Vleck determinant. Observe
\begin{eqnarray}
\Delta_\gamma(s)^{-1}
&=& \det( s^{-1} A ),
\nonumber\\
&=& \exp\hbox{\rm tr}\ln( s^{-1} (I - G_R Q)^{-1} s  ),
\nonumber\\
&=& \exp\hbox{\rm tr}
\ln\left( I + s^{-1}[G_R Q]s  + s^{-1}[G_R Q]^2 s +\cdots\right),
\nonumber\\
&=& \exp\hbox{\rm tr}\left(  s^{-1}[G_R Q ]s  + O(Q^2) \right)_.
\end{eqnarray}
Now, recall that the determinant and trace in these formulae are
to be taken in the sense of $(d+1)\times(d+1)$ matrices, so in terms of
the Ricci tensor
\begin{equation}
\hbox{\rm tr}[Q] =  - \left( R_{\alpha\beta} t^\alpha t^\beta \right)_.
\end{equation}
The $\hbox{\rm tr}[O(Q^2)]$ terms involve messy contractions depending
quadratically on the Riemann tensor. In a weak field approximation
one may neglect these higher order terms and write
\begin{equation}
\Delta_\gamma(s)
= \exp\left(  -s^{-1}(G_R \hbox{\rm tr}[Q])s  + O(Q^2) \right)_.
\end{equation}
More explicitly
\begin{equation}
\Delta_\gamma(x,y)
= \exp\left(  {1\over s}
            \int_0^s (s-s')
            \left( R_{\alpha\beta} t^\alpha t^\beta \right)
             s' ds'
             + O([\hbox{\rm Riemann}]^2) \right)_.
\end{equation}
This approximation, though valid only for weak fields, has a very
nice physical interpretation. The implications of constraints such
as the (timelike/ null/ spacelike) convergence conditions can now be
read off by inspection.

Another nice feature of the weak--field approximation is that the
results for null geodesics can also be read off by inspection
\begin{equation}
\Delta_\gamma(x,y)
= \exp\left(  {1\over \zeta}
            \int_0^\zeta (\zeta-\zeta')
            \left( R_{\alpha\beta} l^\alpha l^\beta \right)
             \zeta' d\zeta'
             + O([\hbox{\rm Riemann}]^2) \right)_.
\end{equation}
Consider, for example, the effect of a thin perfect fluid in an
almost flat background geometry. The Ricci tensor is related to
the density and pressure via the Einstein equations, with the result
that for a null geodesic
\begin{equation}
\Delta_\gamma(x,y) \approx
 \exp\left( 8\pi G
            \int_0^\zeta [\rho + p]
	    {(\zeta-\zeta')  \zeta' \over \zeta}
	    d\zeta' \right)_.
\end{equation}
Unsurprisingly, one sees that energy density and/or pressure located
at the half--way point of the null geodesic is most effective in
terms of focussing.

\subsection{Short distance approximation}

Another result readily derivable from the formal solution (\ref{formal})
is a short distance approximation. Assume that $x$ and $y$ are
close to each other, so that $s_{\gamma}(x,y)$ is small. Assume
that the Riemann tensor does not fluctuate wildly along the geodesic
$\gamma$. Then one may approximate the Ricci tensor by a constant,
and explicitly perform the integration over arc length, keeping
only the lowest order term in $s_{\gamma}(x,y)$
\begin{equation}
\Delta_{\gamma}(x,y)
= 1 + {1\over6}
            \left( R_{\alpha\beta} t^\alpha t^\beta \right)
             s_{\gamma}(x,y)^2
             + O(s_{\gamma}(x,y)^3).
\end{equation}
This result can also be derived via a veritable orgy of index
gymnastics and point--splitting techniques \cite{deWitt-I}, see
equation (1.76) on page 233.

By adopting Gaussian normal coordinates at $x$ one may write \hfil\break
$s_{\gamma}(x,y) \;  t_\mu(x;\gamma; x \leftarrow y) = (x-y)^\mu
+ O[s^2]$ to yield
\begin{equation}
\Delta_{\gamma}(x,y)
= 1 + {1\over6}
            \left( R_{\alpha\beta} (x-y)^\alpha (x-y)^\beta \right)
             + O(s_{\gamma}(x,y)^3).
\end{equation}
In this form the result is blatantly applicable to null geodesics
without further ado.

\subsection{Asymptotics}

To proceed beyond the weak field approximation, consider an
arbitrarily strong gravitational source in an asymptotically flat
spacetime.  Let the point $y$ lie anywhere in the spacetime, possibly
deep within the strongly gravitating region.  Consider an otherwise
arbitrary spacelike or timelike geodesic that reaches and remains
in the asymptotically flat region.

Far from the source the metric is approximately flat and the Riemann
tensor is of order $O[M/r^3]$.  Using the tidal evolution equation,
a double integration with respect to arc length provides the estimate
\begin{equation}
A^\mu{}_\nu(s) =
(A_0)^\mu{}_\nu + s (B_0)^\mu{}_\nu + O[M/s].
\end{equation}
This estimate being valid for large values of $s$, where one is in
the asymptotically flat region. Here $(A_0)$ and $(B_0)$ are constants
that effectively summarize gross features of the otherwise messy
strongly interacting region.  For the van Vleck determinant
$\Delta_\gamma(s)^{-1} = \det\{A/s\} = \det\{B_0 + (A_0/s) +
O[1/s^2]\}$.  Thus the van Vleck determinant approaches a finite
limit at spatial or temporal infinity, with $\Delta_\gamma(\infty)
= \det\{B_0\}^{-1}$.  Finally, define $J = (B_0)^{-1} (A_0)$, to
obtain
\begin{equation}
\Delta_\gamma(s) = {\Delta_\gamma(\infty)\over
\det\{I + (J/s) + O[1/s^2]\} }.
\end{equation}
Which is our desired asymptotic estimate of the van Vleck determinant.

\section{DISCUSSION}

This paper has presented a number of formal developments with regard
to evaluating the van Vleck determinant.  The evolution of the van
Vleck determinant as one moves along a geodesic has been studied
in some detail. By utilizing the Raychaudhuri equation stringent
limits have been placed on this evolution. There is some hope that
it might be possible to formulate more general theorems requiring
weaker convergence hypotheses than those discussed in this paper.
The evolution of the van Vleck determinant has also been reformulated
in terms of tidal effects. Such a presentation has several technical
advantages. Amoung other things, it makes manifest the influence
of the full Riemann tensor.  A formal solution to the resulting
integral equation was presented.  This formal solution in terms of
an iterated integral equation has the very nice property of
terminating in a finite number of steps if the curvature is confined
to thin shells.  Using these formal techniques, a weak field
approximation was also developed. This weak field approximation
depends only on the Ricci tensor.  In a similar vein, a short
distance approximation was developed. Finally, the asymptotic
behaviour at infinity was investigated. The mathematical machinery
developed in this paper has a wide arena of applicability.
Accordingly, some effort has been made to keep  the discussion as
general as reasonably possible.

\underline{Acknowledgements:}

I wish to thank Carl Bender for useful discussions.  I also wish
to thank the Aspen Center for Physics for its hospitality.  This
research was supported by the U.S. Department of Energy.


\end{document}